\documentclass[aps,showpacs,prl,twocolumn,amsmath,amssymb,superscriptaddress]{revtex4}

\usepackage{epsfig}

\begin{document}

\title{Tunneling Density of States, Pair Correlation, and Josephson Current
in Spin-Incoherent Luttinger Liquid-Superconductor Hybrid Systems}

\author{Dagim Tilahun}
\affiliation{Department of Physics, The University of Texas at
Austin, Austin, TX 78712  USA} \affiliation{Kavli Institute for
Theoretical Physics, University of California, Santa Barbara, CA
93106 USA}
\author{Gregory A. Fiete}
\affiliation{Kavli Institute for Theoretical Physics, University
of California, Santa Barbara, CA 93106 USA}\affiliation{Department of
Physics, California Institute of Technology, MC 114-36, Pasadena,
CA 91125 USA}

\date{\today}

\begin{abstract}

We study a hybrid system consisting of a spin-incoherent Luttinger liquid adjoined at one or both ends to a superconductor. We find the
tunneling density of states diverges at low energies and exhibits a universal frequency dependence independent of the strength of the
interactions in the system. We show that in spite of exponentially decaying pair correlations with distance into the spin-incoherent Luttinger
liquid, the Josephson current remains robust. Compared to the zero temperature Luttinger liquid case there is a factor of 2 reduction in the
critical current and a halving of the period in the phase difference between the superconductors. Our results open the way for a new class of
experiments in the spin-incoherent regime of one dimensional systems.

\end{abstract}



\maketitle

The low energy behavior of gapless one dimensional systems, which may be realized by electrons in quantum wires or nanotubes, is described by
Luttinger liquid (LL) theory \cite{Giamarchi}. The elementary excitations of the interacting theory are decoupled bosonic charge and spin modes
that propagate with different velocities, a phenomenon known as \emph{spin-charge separation} that has already been observed experimentally
\cite{Auslaender:sci05}. Strong repulsive interactions tend to suppress the spin velocity while enhancing the charge one, thereby accentuating
the spin-charge separation. For strong enough interactions a window of energy opens at finite temperature where the spin sector consists of
thermally excited (randomized) states  while the charge sector is essentially at zero temperature. A one-dimensional (1-d) system in this regime
is known as a spin-incoherent Luttinger liquid (SILL) \cite{Fiete:rmp_pre}.

While the theory of the SILL has progressed rapidly \cite{cheianov03}, the challenge of reaching the desired window of energy has slowed
experiment.  To date, the best experimental evidence has appeared in momentum resolved tunneling on gated quantum wires \cite{Steinberg:prb06}.
Unfortunately, the analysis of the experiments is somewhat involved \cite{Fiete:rmp_pre} and it has become highly desirable to propose (and
carry out) new experiments to probe the SILL.  Recent experimental progress has made it possible to fabricate devices consisting of nanotubes
\cite{Kasumov:prb03} or quantum wires \cite{Xiang:natn06} between two superconductors (SC).  Through gating to modulate the electron density and
interaction strength, such devices open the possibility of studying the SILL in SILL-SC hybrid structures.

In this Letter we address the theory of  SILL-SC hybrid structures. Various aspects of LL-SC structures have been discussed in the literature
already, including the tunneling density of states \cite{Winkelholz:prl96}, pair correlations \cite{Maslov:prb94}, and Josephson current
\cite{Maslov:prb94,Takane:jpsj97}. Compared to the LL case, the SILL-SC structures exhibit a number of remarkable features. In particular, we
find that the tunneling density of states of a SILL contacted to a SC diverges at low frequencies with a universal form independent of the
strength of the interactions in the system. By contrast, for the isolated LL, SILL, and the LL-SC system the energy dependence of the tunneling
density of states depends on the strength of the interactions.  We also compute the decay of the pair correlations into the SILL from the SC
and, as might be expected from the highly excited spin states, they decay exponentially fast with a length scale set by the interparticle
spacing.  However, the Josephson current resulting from the {\em coherent} propagation of Cooper pairs through a finite length SILL remains
robust.  Moreover, the critical current shows the same scaling with length as in the case of a LL but its value is reduced by a factor of two.
As a function of phase between the two superconductors, the period of the Josephson current is halved.  Both the tunneling density of states and
the Josephson current should be experimentally accessible. Observation of the results described here would be a smoking gun signature of the
SILL in these hybrid structures.

\begin{figure}[h]
\includegraphics[width=.85\linewidth,clip=]{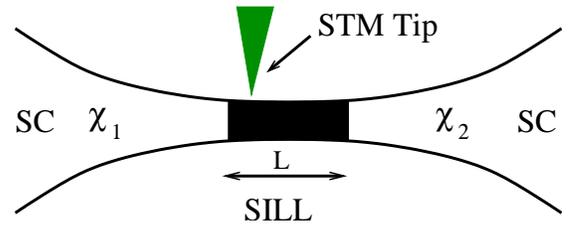}
 \caption{\label{fig:schematic} (color online) Schematic of the model we study.  A spin-incoherent Luttinger liquid (SILL) of length $L$ is
 adiabatically connected to two superconductors (SC) with phase $\chi_1$ and $\chi_2$, and the same superconducting gap $\Delta$.
 The tunneling density of states a distance $x$ from the end of the SILL can be probed with electron tunneling from a metallic lead
 such as scanning tunneling microscope (STM) tip. A Josephson current flows through the SILL when $\chi\equiv \chi_1-\chi_2 \neq 0$.}
 \end{figure}

A schematic of our model is shown in Fig.1.  We assume the SILL is adiabatically connected to the SC so that the scattering at the interface is
in the Andreev limit.  In all our calculations we assume that $k_B T \ll \Delta$ where $k_B$ is Boltzmann's constant, $T$ is the temperature,
and $\Delta$ is the magnitude of the superconducting gap. Bosonization procedures have been developed for LL-SC hybrid systems
\cite{Maslov:prb94,Takane:jpsj96} valid for energies $\omega \ll \Delta \ll E_F$ that are also small compared to the characteristic spin and
charge energies. At the lowest energy scales the Hamiltonian is given by $H = H_\rho + H_\sigma$ where
\begin{equation}
H_\rho  = v_\rho\int \frac{dx}{2\pi}
\left[\frac{1}{g_\rho}(\partial_x \theta_\rho(x))^2 +
g_\rho(\partial_x \phi_\rho(x))^2\right],\label{chargehamltn}
\end{equation}
and $H_\sigma$ has the same form only with $\rho$ replaced by $\sigma$. Here $\theta_\rho(x)$ and $\phi_\rho(x)$ are bosonic fields representing
charge and current density fluctuations, $g_\rho$ measures the strength of the interactions ($g_\rho = 1$ for non-interacting systems, $g_\rho <
1$ for repulsive interactions, and $g_\sigma=1$ for $SU(2)$ invariant spin interactions), and $v_\rho$ is the velocity of the charge modes. The
mode expansions of the bosonic fields for a LL between two superconductors separated by a distance $L$ are \cite{Maslov:prb94}
\begin{eqnarray}
\theta_\rho(x) &=& \theta^{(o)}_\rho + \sqrt{g_\rho}\sum_{q>0}\gamma_q\cos(qx)(b^\dagger_{\rho q}+b_{\rho q}),\nonumber\\
\theta_\sigma(x) &=& \frac{\pi}{\sqrt{2}}M\frac{x}{L} +\sqrt{g_\sigma}\sum_{q>0}\gamma_q\sin(qx)(b^\dagger_{\sigma q}-b_{\sigma q}),\nonumber\\
\phi_\rho(x) &=& \frac{\pi}{\sqrt{2}}\left(J'+\frac{\chi}{\pi}\right)\frac{x}{L}
+\frac{1}{\sqrt{g_\rho}}\sum_{q>0}\gamma_q\sin(qx)(b^\dagger_{\rho q}
-b_{\rho q}),\nonumber\\
\phi_\sigma(x) &=& \phi^{(o)}_\sigma +\frac{1}{\sqrt{g_\sigma}}\sum_{q>0}\gamma_q\cos(qx)(b^\dagger_{\sigma q}+b_{\sigma
q}),\label{modexpansion}
\end{eqnarray}
where $\gamma_q = \sqrt{\frac{\pi}{qL}}e^{-\alpha q}$ with $\alpha$ a short distance cutoff on the scale of the interparticle spacing,
$[b_{\beta q},b^\dagger_{\beta', q'}]=\delta_{\beta \beta'} \delta_{qq'}$ for $\beta=\rho,\sigma$, and $\chi$ is the phase difference of the
order parameter of the two superconductors (assumed to have the same $\Delta$). The integers $J'$ and $M$ are the topological (zero mode)
numbers \cite{haldane} that are related to excess charge and spin densities. They obey the constraint $J'+ M = $ even.  Substituting the
expansion \eqref{modexpansion} into \eqref{chargehamltn} leads to
\begin{eqnarray}
H&=&\frac{\pi}{4L}\left(v_\rho g_\rho\left(J'+\frac{\chi}{\pi}\right)^2+\frac{v_\sigma}{g_\sigma}M^2\right)\nonumber\\&+&\sum_{q>0}
q\left(v_\rho a^{\dagger}_{\rho q}a_{\rho q}+v_\sigma a^{\dagger}_{\sigma q}a_{\sigma q}\right).\label{diagenrgy}
\end{eqnarray}
While the form of the energy contribution from the non-zero modes ($q>0$) is valid only for low energies relative to the spin and charge energy,
the zero mode contribution is valid at all energies, in particular in the spin-incoherent regime defined by the condition $E_{\rm spin} \ll k_B T \ll E_{\rm charge}$ where $E_{\rm spin/charge}=\frac{v_{\sigma/\rho}}{L}$ for zero mode properties and $E_{\rm spin/charge}=\frac{v_{\sigma/\rho}}{\alpha}$ for fluctuating quantities like the Greens function and pair correlations. In terms of the fields \eqref{modexpansion} the bosonized electron annihilation operator is $\psi_s(x,t)\sim
\frac{1}{\sqrt{\alpha L}}e^{i(\theta_s(x,t)-\phi_s(x,t))}+\frac{1}{\sqrt{\alpha L}}e^{-i(\theta_s(x,t)+\phi_s(x,t))}$ where
$s=\uparrow,\downarrow$.  The charge fields $\theta_\rho=(\theta_\uparrow+\theta_\downarrow)/\sqrt{2}$ and the spin fields
$\theta_\sigma=(\theta_\uparrow-\theta_\downarrow)/\sqrt{2}$ with identical definitions for $\phi_\rho$ and $\phi_\sigma$.

{\em Single particle Greens function.}  We compute the single particle Green's function $G_s(x,x';\tau,\tau ') = -\langle
T_\tau\psi_s(x,\tau)\psi^{\dagger}_s(x',\tau')\rangle$  for a SILL connected to a {\em single} SC.  Fiete and Balents \cite{Fiete:prl04} have
developed a simple but powerful method to evaluate such expectation values and we employ it here. Let us first consider the trace over the spin
sector, which we assume consists of highly thermalized random spins. The dominant contribution comes from the terms where there are effectively
no exchanges of particles for all the particles between $(x',\tau ')$ and $(x,\tau)$.  For the single particle Greens function, this implies
that all spins have same orientation between $(x',\tau ')$ and $(x,\tau)$ and this occurs with a probability of $2^{-|N(x,\tau;x',\tau ')|}$
where $N(x,\tau;x',\tau ')$ is the number of electrons between the two points. A factor $(-1)^{N(x,\tau;x',\tau ')}$ arises from the permutation
of the propagating electron with the other electrons in the SILL. The general result is \cite{Fiete:prl04}
\begin{eqnarray}
G_\sigma(x,x';\tau,\tau ') \sim \sum_{m=-\infty}^{\infty}\langle
\delta\left(m-N(x,\tau;x',\tau ')\right)(-1)^m\nonumber\\\times
2^{-|m|}
e^{-\frac{i}{\sqrt{2}}(\phi_\rho(x,\tau)-\phi_\rho(x',\tau
'))}\rangle,\label{grnfn}
\end{eqnarray}
where the remaining expectation value is taken over the charge degrees of freedom at zero temperature. The number operator is related to the
$\theta_\rho$ field,
\begin{equation}
\label{eq:N}
N(x,\tau;x',\tau ') = \bar n(x-x') +
\frac{\sqrt{2}}{\pi}(\theta_\rho(x,\tau)-\theta_\rho(x',\tau ')),
\end{equation}
with $\bar n = \frac{1}{a}= 2k_F/\pi$, where $k_F$ is the Fermi wavevector and $a$ is the mean interparticle spacing. For a system with a
boundary, $N$ and $G_s$ are not space translationally invariant \cite{Kindermann_crossover:prb06}.

When $|x-x'|$ is small, the Green function can be expressed as \cite{Fiete:rmp_pre},
\begin{eqnarray}
G_s(x,x';\tau,\tau ') \sim \sqrt{\frac{\pi}{\langle \Theta_\rho^2
\rangle}}\sum_{m=-\infty}^\infty 2^{-|m|}(-1)^m\nonumber\\\times
e^{-\frac{\pi^2(\bar
n(x-x')-m)^2}{\langle\Theta_\rho^2\rangle}}e^{-\frac{\langle
\Phi_\rho^2\rangle}{4}},\label{grnfnsmllx}
\end{eqnarray}
where $\Theta_\rho$ and $\Phi_\rho$ are
$\Theta_\rho(x,\tau;x',\tau ')\equiv \theta_\rho
(x,\tau)-\theta_\rho (x',\tau ')$ and $\Phi_\rho(x,\tau;x',\tau
')\equiv \phi_\rho (x,\tau)-\phi_\rho (x',\tau')$, respectively. For $x=x'$ and
$\tau ' = 0$, the expansions \eqref{modexpansion} with $L\to \infty$ give
$
\langle \Phi_\rho^2 \rangle =
\frac{1}{2g_\rho}\left[2\ln\left(\frac{2x}{\alpha}\right)+\ln\left(\frac{(v_\rho
\tau)^2}{(v_\rho \tau)^2+(2x)^2}\right)\right],
$
and
$
\langle \Theta_\rho^2 \rangle =\frac{g_\rho}{2}\left[4\ln\left(\frac{v_\rho\tau}{\alpha}\right)-
2\ln\left(\frac{2x}{\alpha}\right)-\ln\left(\frac{(v_\rho
\tau)^2}{(v_\rho \tau)^2+(2x)^2}\right)\right].
$
The resulting Green's function (valid for $x,v_\rho \tau > \alpha$) is
\begin{widetext}
\begin{equation}
G_s(x,x;\tau,0)\sim
\frac{1}{\sqrt{g_\rho\left(\ln\left(\frac{(v_\rho \tau)^2}{\alpha^2}\right)
-\frac{1}{2}\ln\left(\frac{(2x)^2}{\alpha^2} \frac{(v_\rho \tau)^2}{(2x)^2+
((v_\rho \tau)^2}\right)\right)}}\left(\frac{\alpha^2}{(2x)^2}\right)^{\frac{1}{8g_\rho}}
\left(\frac{(v_\rho \tau)^2}{(2x)^2+((v_\rho \tau)^2}\right)^{-\frac{1}{8g_\rho}} ,
\end{equation}
\end{widetext}
where we have kept only the dominant $m=0$ term \cite{Fiete:prl04}.

{\em Tunneling Density of States.} The local tunneling density of states, $A(x,\omega)$, can be computed by Fourier transforming the Greens
function.  For $v_\rho \tau \gg x$, the Green's function $G_s(x,x;\tau,0)\sim
\left(\frac{\alpha^2}{(2x)^2}\right)^{\frac{1}{8g_\rho}}\frac{1}{\sqrt{g_\rho\ln(v_\rho \tau/\alpha)}}$ which implies
\begin{equation}
A_{\rm SILL-SC}(x,\omega)\sim \left(\frac{\alpha^2}{(2x)^2}\right)^{\frac{1}{8g_\rho}} \frac{\omega^{-1}}{\sqrt{|\ln(\omega)|}},\;\; \omega>0.
\label{DOS}
\end{equation}
Thus, the tunneling density of states of a SILL-SC hybrid system diverges at low energies in a universal manner independent of the interaction
strength $g_\rho$. The result \eqref{DOS} is quite dramatic in light of known results for other related systems.  If a LL with $SU(2)$ symmetric
interactions is attached to a SC,  $A_{\rm LL-SC}(x,\omega) \sim  \left(\frac{\alpha^2}{(2x)^2}\right)^{\frac{1}{8g_\rho}-\frac{g_\rho}{8}}
\omega^{\frac{1}{2}(g_\rho-1)}$ which also diverges for repulsive interactions ($g_\rho <1$) but in an interaction dependent way
\cite{Winkelholz:prl96}.  By studying the frequency dependence of the tunneling density of states for different distances from the boundary
(with a scanning tunneling microscope, for example) both $g_\rho$ and whether or not the 1-d system is in the LL or SILL regime can be deduced.
By fixing a position $x$ and changing the density of the wire with a back gate as in Ref.\cite{Auslaender:sci05}, $g_\rho$ can be tuned allowing
a test of the interaction dependence (or lack thereof) of the low frequency behavior of  $A(x,\omega)$ thus enabling an unambiguous
determination of SILL physics.


{\em Pair Correlations.} We now turn our attention to the pair correlation, $F(x) \equiv -\langle
\psi_{\uparrow}(x)\psi_{\downarrow}(x)\rangle$, the most natural measure of the proximity effect.   We assume the Cooper pairs leak into the
SILL at $x'=0$, and calculate their amplitude at distance $x> \alpha$.  The calculation in the spin-incoherent case closely follows that of the
Green's function \cite{Fiete:rmp_pre,Fiete:prl04}.  The no particle exchange condition here tells us that the dominant contribution from the
spin sector is an alternating up/down configuration.  The probability of finding an up/down configuration from the boundary to a distance $x$ is
$2^{-|N(x,0;0,0)|}$ where $N$ is given by \eqref{eq:N}. Fermi statistics gives a factor  $(-1)^{2N(x)}$ which trivially evaluates to unity
(because the Cooper pair is effectively a boson), and the annihilation of the two electrons is accomplished by the operator
$e^{-i2\phi_\rho(x)/\sqrt 2}$.  Combining these after taking the spin trace gives
\begin{eqnarray}
F(x) \sim e^{-2k_F |x|\frac{\ln2}{\pi}}\langle e^{-2 \frac{\ln2}{\pi} \Theta_\rho(x,0;0,0)/{\sqrt 2}} e^{-i 2 \phi_\rho(x)/\sqrt 2}\rangle \nonumber\\
\sim e^{-2k_F |x|\frac{\ln2}{\pi}} \left(\frac{\alpha}{x}\right)^{\frac{1}{2g_\rho}-\frac{3g_\rho}{2}\left(\frac{\ln2}{\pi}\right)^2},
\label{pairamp}
\end{eqnarray}
so that the pair correlations decay exponentially fast into the SILL.  Note that the exponential factor is the same as that found for the
single-particle Greens function in an isolated SILL \cite{Fiete:prl04}.  This is, in fact, a general result in the spin-incoherent regime: all
$n$-body particle non-conserving operators decay with the same exponential factor because the no exchange condition is a result independent of
$n$ (for sufficiently large $x$).  It is instructive to compare the result \eqref{pairamp} with that obtained \cite{Maslov:prb94,Fazio:sm99} for
a LL when $T\lesssim v_\sigma/a \ll v_\rho/a$ using the mode expansions \eqref{modexpansion}: $F(x)\sim e^{-\frac{g_\sigma|x|}{2\xi(T)}}
\left(\frac{\alpha}{x}\right)^{\frac{1}{2g_\rho}} \left(\frac{\alpha}{\xi(T)}\right)^{\frac{g_\sigma}{2}}$ where the spin correlation length
$\xi(T)=v_\sigma/(\pi k_B T)$.  As discussed in Ref.\cite{Fiete:prb05} the SILL can often be understood from the LL result when  $\xi(T)$
saturates to a number of order the mean particle spacing: $\xi(T)={\cal O}(a)$.  Clearly this is also the case for the pair correlations, aside
from the correction $-\frac{3g_\rho}{2}\left(\frac{\ln2}{\pi}\right)^2$ to the exponent in \eqref{pairamp} from Gaussian density fluctuations
(reminiscent of the Greens function \cite{Fiete:prl04}).

\begin{figure}[h]
\includegraphics[width=.95\linewidth,clip=]{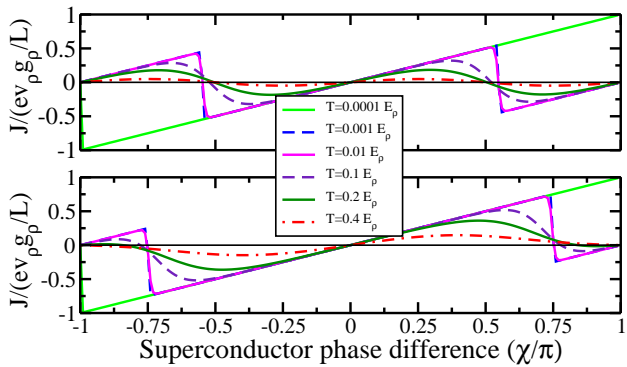}
 \caption{\label{fig:Josephson} (color online) Josephson current as a function of phase difference $\chi$ between two superconductors
 and temperature in units of $E_\rho=\pi v_\rho g_\rho/(4L)=\epsilon_\rho (k_B T)$.  Top:  $E_\sigma/E_\rho=0.1$.
 Bottom: $E_\sigma/E_\rho=0.5$.  Here $E_\sigma=\pi v_\sigma/(g_\sigma 4 L)=\epsilon_\sigma (k_B T).$ }
 \end{figure}

{\em Josephson Current.} It is a remarkable result that a current of Cooper pairs flows between two superconductors when the phase difference
$\chi$ between them is non-zero (See Fig. \ref{fig:schematic}). The Josephson current may be obtained from the well known thermodynamic relation
\cite{Maslov:prb94,Takane:jpsj97}
\begin{equation}
J(\chi) =-2ek_B T\frac{\partial \ln Z}{\partial \chi},
\label{jsphformula}
\end{equation}
where $Z$ is the partition function of the 1-d system and $-e$ is the charge of the electron.  The LL case has been investigated before
\cite{Maslov:prb94,Takane:jpsj97}. Our approach here is to assume our system is initially in the LL regime at $T=0$ and then take the
$v_\sigma/L \ll k_B T \ll v_\rho/L$ temperature limit \cite{Fiete:prb05} using the mode expansions \eqref{modexpansion} whose zero mode
components are valid at all energies, in particular in the spin-incoherent regime.  At all energies the ``$J'+M=$ even" constraint relates the
charge and spin parts of the topological (zero mode) terms \cite{Maslov:prb94} of the Hamiltonian \eqref{diagenrgy}.  Since the topological
terms are decoupled from the non-zero modes, the partition function factorizes, $Z=Z(\chi)\widetilde{Z}$, where
$Z(\chi)=\sum_{J'+M=even}e^{-\epsilon_\rho\left(J'+\frac{\chi}{\pi}\right)^2-\epsilon_\sigma M^2}$ is the partition function of the zero modes
with $\epsilon_\rho = \frac{\pi v_\rho g_\rho}{4Lk_B T}$ and $\epsilon_\sigma = \frac{\pi v_\sigma}{4g_\sigma Lk_BT}$ and $\widetilde{Z}$
describes the non-zero modes. To remove the ``$J'+M=$ even"  constraint, we let $J' = 2j+j_i$ and $M = 2m + m_i$, with $j,m = 0, \pm 1, \dots$
and sum over the two sectors  $j_i=m_i=0,1$. Thus, $Z(\chi) = \sum_{j,m}e^{-\epsilon_{\rho}(2j+\frac{\chi}{\pi})^2}e^{-\epsilon_\sigma
(2m)^2}+e^{-\epsilon_{\rho}(2j+1+\frac{\chi}{\pi})^2}e^{-\epsilon_\sigma (2m+1)^2}.$ Note $Z(\chi +2\pi) = Z(\chi)$; thus we restrict our domain
to $|\chi| \le \pi$.  For low temperatures and high temperatures relative to both charge and spin velocities, $J(\chi)$ has been worked out in
Ref.\cite{Maslov:prb94}.  Here we focus on the spin-incoherent case where $\epsilon_\sigma \ll 1 \ll \epsilon_\rho$.  Our exact evaluation of
$J(\chi)$ is plotted in Fig. \ref{fig:Josephson} for $\epsilon_\sigma/\epsilon_\rho=0.1$ and $\epsilon_\sigma/\epsilon_\rho=0.5$.

At the lowest temperatures the zero temperature result $J(\chi)=\left(\frac{ev_\rho g_\rho}{L}\right) \frac{\chi}{\pi}$ coming from the
$m_i=j_i=0$ terms is well approximated in both cases. However, at slightly higher temperatures contributions appear from the $m_i=j_i=1$ terms
in the partition function and create an additional zero in $J(\chi)$ for $\pi/2 \leq |\chi| < \pi$.  The location of this zero depends on
temperature, but for temperatures low compared to $E_\rho$ the zero is most sensitive to the ratio of spin to charge energy, $E_\sigma/E_\rho$.
Thus, the zeros of $J(\chi)$ may be used to deduce the ratio $E_\sigma/E_\rho$.  In the limit $E_\sigma/E_\rho\ll 1$ this zero moves to $\pm
\pi/2$.  As Fig. \ref{fig:Josephson} shows, already for $E_\sigma/E_\rho=0.1$ the limit $E_\sigma/E_\rho\ll 1$ is approached.  The condition
$E_\sigma/E_\rho\ll 1$ is a prerequisite for SILL physics, so $J(\chi)$ can establish whether the 1-d system has large enough spin-charge
separation for SILL physics.  In the extreme limit $E_\rho \gg k_B T \gg E_\sigma \to0$, the exact form of the spin-incoherent Josephson current
can be obtained analytically:
\begin{equation}
\frac{J(\chi)}{\left(\frac{ev_\rho g_\rho}{L}\right)} = \Biggl \{\begin{array}{cc} \frac{\chi}{\pi}+1,& -\pi < \chi <- \frac{\pi}{2}\\
\frac{\chi}{\pi},& -\frac{\pi}{2} < \chi <
\frac{\pi}{2}\\\frac{\chi}{\pi}-1, & \frac{\pi}{2} < \chi < \pi
\end{array}. \label{joscrntSILL}
\end{equation}
Compared to the LL case at $T=0$, both the period and the critical current are halved in the SILL regime. The latter is reminiscent of Matveev's
result for the conductance of a quantum wire in the SILL regime adiabatically connected to Fermi liquid leads, where the conductance reduces to
$\frac{e^2}{h}$ per mode rather than the $T=0$ universal value of $\frac{2e^2}{h}$ \cite{Matveev:prl04}.  Thus, the observation of a Josephson
current that follows \eqref{joscrntSILL} is a clear indication of SILL behavior.  It is remarkable that in spite of the exponentially decaying
pair correlations \eqref{pairamp} the Josephson current remains robust, only with a factor of two reduction in the critical current compared to
the $T=0$ LL result. Physically, this is because the SC phase difference $\chi$ couples only to the charge degrees of freedom which remain coherent in the SILL. Note also that the critical current in the SILL regime scales as $\sim 1/L$ as general arguments require
\cite{Maslov:prb94}.

In summary, we have determined the properties expected for a SILL adiabatically connected to one or two superconductors.  The tunneling density
of states exhibits a universal frequency dependence independent of interactions in the system and the Josephson current has a saw-tooth form
with a factor of two reduction in the critical current and a halving of the period.  If the contacts of the SILL are non-ideal (non-adiabatic),
the adiabatic regime may still be obtained at low energies as impurities are irrelevant in the SILL for $g_\rho > 1/2$ \cite{Fiete:prb05}. In
this sense, the adiabatic model is even more relevant for the SILL than for the LL.

We are grateful to Oleg Starykh for helpful discussions and for financial support from NSF Grants PHY05-51164,
DMR-0606489, the Lee A. DuBridge Foundation and the Welch Foundation.


\begin{thebibliography}{27}
\expandafter\ifx\csname natexlab\endcsname\relax\def\natexlab#1{#1}\fi
\expandafter\ifx\csname bibnamefont\endcsname\relax
  \def\bibnamefont#1{#1}\fi
\expandafter\ifx\csname bibfnamefont\endcsname\relax
  \def\bibfnamefont#1{#1}\fi
\expandafter\ifx\csname citenamefont\endcsname\relax
  \def\citenamefont#1{#1}\fi
\expandafter\ifx\csname url\endcsname\relax
  \def\url#1{\texttt{#1}}\fi
\expandafter\ifx\csname urlprefix\endcsname\relax\def\urlprefix{URL }\fi
\providecommand{\bibinfo}[2]{#2}
\providecommand{\eprint}[2][]{\url{#2}}

\bibitem[{\citenamefont{Giamarchi}(2004)}]{Giamarchi}
\bibinfo{author}{\bibfnamefont{T.}~\bibnamefont{Giamarchi}},
  \emph{\bibinfo{title}{Quantum Physics in One Dimension}}
  (\bibinfo{publisher}{Clarendon Press, Oxford}, \bibinfo{year}{2004}).

\bibitem[{\citenamefont{Auslaender et~al.}(2005)\citenamefont{Auslaender,
  Steinberg, Yacoby, Tserkovnyak, Halperin, Baldwin, Pfeiffer, and
  West}}]{Auslaender:sci05}
\bibinfo{author}{\bibfnamefont{O.~M.} \bibnamefont{Auslaender {\it et al.}}},
  \bibinfo{journal}{Science} \textbf{\bibinfo{volume}{308}},
  \bibinfo{pages}{88} (\bibinfo{year}{2005}).


\bibitem[{\citenamefont{Fie}(2007)}]{Fiete:rmp_pre}
\bibinfo{author}{\bibfnamefont{G.~A.} \bibnamefont{Fiete}},
  \bibinfo{journal}{Rev. Mod. Phys.} \textbf{\bibinfo{volume}{79}},
  \bibinfo{pages}{801} (\bibinfo{year}{2007}).

\bibitem[{\citenamefont{Cheianov and Zvonarev}(2004)}]{cheianov03}
\bibinfo{author}{\bibfnamefont{V.~V.} \bibnamefont{Cheianov}} \bibnamefont{and}
  \bibinfo{author}{\bibfnamefont{M.~B.} \bibnamefont{Zvonarev}},
  \bibinfo{journal}{Phys. Rev. Lett.} \textbf{\bibinfo{volume}{92}},
  \bibinfo{pages}{176401} (\bibinfo{year}{2004});
%
\bibinfo{author}{\bibfnamefont{M.}~\bibnamefont{Kindermann}},
  \bibinfo{author}{\bibfnamefont{P.~W.} \bibnamefont{Brouwer}},
  \bibnamefont{and} \bibinfo{author}{\bibfnamefont{A.~J.}
  \bibnamefont{Millis}}, \bibinfo{journal}{{\it ibid.}}
  \textbf{\bibinfo{volume}{97}}, \bibinfo{pages}{036809}
  (\bibinfo{year}{2006});
%
\bibinfo{author}{\bibfnamefont{K.~A.} \bibnamefont{Matveev}},
  \bibinfo{author}{\bibfnamefont{A.}~\bibnamefont{Furusaki}}, \bibnamefont{and}
  \bibinfo{author}{\bibfnamefont{L.~I.} \bibnamefont{Glazman}},
  \bibinfo{journal}{{\it ibid.}} \textbf{\bibinfo{volume}{98}},
  \bibinfo{eid}{096403} (\bibinfo{year}{2007});
%
\bibinfo{author}{\bibfnamefont{O.~F.} \bibnamefont{Sylju\aa sen}},
  \bibinfo{journal}{{\it ibid.}} \textbf{\bibinfo{volume}{98}},
  \bibinfo{eid}{166401} (\bibinfo{year}{2007}).

\bibitem[{\citenamefont{Steinberg et~al.}(2006)\citenamefont{Steinberg,
  Auslaender, Yacoby, Qian, Fiete, Tserkovnyak, Halperin, Baldwin, Pfeiffer,
  and West}}]{Steinberg:prb06}
\bibinfo{author}{\bibfnamefont{H.}~\bibnamefont{Steinberg {\it et al.}}},
  \bibinfo{journal}{Phys. Rev. B} \textbf{\bibinfo{volume}{73}},
  \bibinfo{pages}{113307} (\bibinfo{year}{2006}).

\bibitem[{\citenamefont{Kasumov et~al.}(2003)\citenamefont{Kasumov, Kociak,
  Ferrier, Deblock, Gu\'eron, Reulet, Khodos, St\'ephan, and
  Bouchiat}}]{Kasumov:prb03}
\bibinfo{author}{\bibfnamefont{A.}~\bibnamefont{Kasumov {\it et al.}}},
  \bibinfo{journal}{Phys. Rev. B} \textbf{\bibinfo{volume}{68}},
  \bibinfo{pages}{214521} (\bibinfo{year}{2003});
%
\bibinfo{note}{P.E. Lindeloff,
  http://meetings.aps.org/link/ BAPS.2007.MAR.D1.4.}

\bibitem[{\citenamefont{Xiang et~al.}(2006)\citenamefont{Xiang, Vidan, Tinkham,
  Westervelt, and Lieber}}]{Xiang:natn06}
\bibinfo{author}{\bibfnamefont{J.}~\bibnamefont{Xiang {\it et al.}}},
  \bibinfo{journal}{Nature Nano.}
  \textbf{\bibinfo{volume}{1}}, \bibinfo{pages}{208} (\bibinfo{year}{2006});
%
\bibinfo{author}{\bibfnamefont{Y.-J.} \bibnamefont{Doh {\it et al.}}},
  \bibinfo{journal}{Science} \textbf{\bibinfo{volume}{309}},
  \bibinfo{pages}{272} (\bibinfo{year}{2005}).

\bibitem[{\citenamefont{Winkelholz et~al.}(1996)\citenamefont{Winkelholz,
  Fazio, Hekking, and Sch\"on}}]{Winkelholz:prl96}
\bibinfo{author}{\bibfnamefont{C.}~\bibnamefont{Winkelholz {\it et al.}}},
  \bibinfo{journal}{Phys. Rev. Lett.} \textbf{\bibinfo{volume}{77}},
  \bibinfo{pages}{3200} (\bibinfo{year}{1996}).

\bibitem[{\citenamefont{Maslov et~al.}(1996)\citenamefont{Maslov, Stone,
  Goldbart, and Loss}}]{Maslov:prb94}
\bibinfo{author}{\bibfnamefont{D.~L.} \bibnamefont{Maslov {\it et al.}}},
  \bibinfo{journal}{Phys. Rev. B} \textbf{\bibinfo{volume}{53}},
  \bibinfo{pages}{1548} (\bibinfo{year}{1996}).

\bibitem[{\citenamefont{Takane}(1997)}]{Takane:jpsj97}
\bibinfo{author}{\bibfnamefont{R.}~\bibnamefont{Fazio}},
  \bibinfo{author}{\bibfnamefont{F.~W.~J.} \bibnamefont{Hekking}}, \bibnamefont{and}
  \bibinfo{author}{\bibfnamefont{A.~A.} \bibnamefont{Odintsov}},
  \bibinfo{journal}{Phys. Rev. Lett.} \textbf{\bibinfo{volume}{74}},
  \bibinfo{pages}{1843} (\bibinfo{year}{1995});
  \bibinfo{author}{\bibfnamefont{Y.}~\bibnamefont{Takane}},
  \bibinfo{journal}{J. Phys. Soc. Japan} \textbf{\bibinfo{volume}{66}},
  \bibinfo{pages}{537}
  (\bibinfo{year}{1997});
%
\bibinfo{author}{\bibfnamefont{I.}~\bibnamefont{Affleck}},
  \bibinfo{author}{\bibfnamefont{J.-S.} \bibnamefont{Caux}}, \bibnamefont{and}
  \bibinfo{author}{\bibfnamefont{A.~M.} \bibnamefont{Zagoskin}},
  \bibinfo{journal}{Phys. Rev. B} \textbf{\bibinfo{volume}{62}},
  \bibinfo{pages}{1433} (\bibinfo{year}{2000});
  \bibinfo{author}{\bibfnamefont{J.-S.} \bibnamefont{Caux}},
  \bibinfo{author}{\bibfnamefont{H.}~\bibnamefont{Saleur}},
\bibnamefont{and}
  \bibinfo{author}{\bibfnamefont{F.}~\bibnamefont{Siano}},
  \bibinfo{journal}{Phys. Rev. Lett.} \textbf{\bibinfo{volume}{88}},
  \bibinfo{pages}{106402} (\bibinfo{year}{2002});
\bibinfo{author}{\bibfnamefont{A.~E.} \bibnamefont{Feiguin}},
  \bibinfo{author}{\bibfnamefont{S.~R.} \bibnamefont{White}}, \bibnamefont{and}
  \bibinfo{author}{\bibfnamefont{D.~J.} \bibnamefont{Scalapino}},
  \bibinfo{journal}{Phys. Rev. B} \textbf{\bibinfo{volume}{75}},
  \bibinfo{eid}{024505} (\bibinfo{year}{2007}).

\bibitem[{\citenamefont{Takane and Koyama}(1996)}]{Takane:jpsj96}
\bibinfo{author}{\bibfnamefont{Y.}~\bibnamefont{Takane}} \bibnamefont{and}
  \bibinfo{author}{\bibfnamefont{Y.}~\bibnamefont{Koyama}},
  \bibinfo{journal}{J. Phys. Soc. Japan} \textbf{\bibinfo{volume}{65}},
  \bibinfo{pages}{3630} (\bibinfo{year}{1996}).

\bibitem[{\citenamefont{Haldane}(1981)}]{haldane}
\bibinfo{author}{\bibfnamefont{F.~D.~M.} \bibnamefont{Haldane}},
  \bibinfo{journal}{J. Phys. C} \textbf{\bibinfo{volume}{14}},
  \bibinfo{pages}{2585} (\bibinfo{year}{1981}).

\bibitem[{\citenamefont{Fiete and Balents}(2004)}]{Fiete:prl04}
\bibinfo{author}{\bibfnamefont{G.~A.} \bibnamefont{Fiete}} \bibnamefont{and}
  \bibinfo{author}{\bibfnamefont{L.}~\bibnamefont{Balents}},
  \bibinfo{journal}{Phys. Rev. Lett.} \textbf{\bibinfo{volume}{93}},
  \bibinfo{pages}{226401} (\bibinfo{year}{2004}).

\bibitem[{\citenamefont{Kindermann and
  Brouwer}(2006)}]{Kindermann_crossover:prb06}
\bibinfo{author}{\bibfnamefont{M.}~\bibnamefont{Kindermann}} \bibnamefont{and}
  \bibinfo{author}{\bibfnamefont{P.~W.} \bibnamefont{Brouwer}},
  \bibinfo{journal}{Phys. Rev. B} \textbf{\bibinfo{volume}{74}},
  \bibinfo{pages}{115121} (\bibinfo{year}{2006}).

\bibitem[{\citenamefont{Fazio et~al.}(1999)\citenamefont{Fazio, Hekking,
  Odintsov, and Raimondi}}]{Fazio:sm99}
\bibinfo{author}{\bibfnamefont{R.}~\bibnamefont{Fazio {\it et al.}}},
  \bibinfo{journal}{Superlat. and Microstrs.}
  \textbf{\bibinfo{volume}{25}}, \bibinfo{pages}{1163} (\bibinfo{year}{1999}).

\bibitem[{\citenamefont{Fiete et~al.}(2005)\citenamefont{Fiete, {Le Hur}, and
  Balents}}]{Fiete:prb05}
\bibinfo{author}{\bibfnamefont{G.~A.} \bibnamefont{Fiete}},
  \bibinfo{author}{\bibfnamefont{K.}~\bibnamefont{{Le Hur}}}, \bibnamefont{and}
  \bibinfo{author}{\bibfnamefont{L.}~\bibnamefont{Balents}},
  \bibinfo{journal}{Phys. Rev. B} \textbf{\bibinfo{volume}{72}},
  \bibinfo{pages}{125416} (\bibinfo{year}{2005}).

\bibitem[{\citenamefont{Matveev}(2004{\natexlab{a}})}]{Matveev:prl04}
\bibinfo{author}{\bibfnamefont{K.~A.} \bibnamefont{Matveev}},
  \bibinfo{journal}{Phys. Rev. Lett.} \textbf{\bibinfo{volume}{92}},
  \bibinfo{pages}{106801} (\bibinfo{year}{2004}{\natexlab{a}});
%
\bibinfo{author}{\bibfnamefont{K.~A.} \bibnamefont{Matveev}},
  \bibinfo{journal}{Phys. Rev. B} \textbf{\bibinfo{volume}{70}},
  \bibinfo{pages}{245319} (\bibinfo{year}{2004}{\natexlab{b}}).

\end{thebibliography}

\end{document}